\documentclass[%
 reprint,
 bibnotes,
 amsmath,amssymb,
 aps,
]{revtex4-2}

\usepackage{color}
\usepackage{graphicx}
\usepackage{epstopdf}
\usepackage{dcolumn}
\usepackage{bm}
\usepackage{hyperref}
\usepackage[english]{babel}
\usepackage{subfigure}
\usepackage[utf8]{inputenc}

\begin{document}
\makeatletter

\preprint{APS/123-QED}

\title{Coulomb Blockade in Angstrom-scale latent ion track channels}

\newcommand{\phy}{School of Physical Science and Technology, Northwestern Polytechnical University, Xi’an, 710072, China}
\newcommand{\iem}{School of Aeronautics and Institute of Extreme Mechanics, Northwestern Polytechnical University, Xi’an, 710072, China}

\author
{Yanbo Xie,$^{1\ast,2}$ Deli Shi,$^{2}$ Wenhui Wang$^{2}$ and Ziheng Wang$^{2}$
\\
\normalsize{$^{1}$School of Aeronautics and Institute of Extreme Mechanics, Northwestern Polytechnical University}\\
\normalsize{$^{2}$School of Physical Science and Technology, Northwestern Polytechnical University}\\
\normalsize{Xi’an, 710072, China}
}

\date{\today}

\begin{abstract}
When channels were scaled down to the size of hydrated ions, ionic Coulomb blockade was discovered. However, the experimental CB phenomenon was rarely reported since Feng et.al., discovered in $MoS_2$ nanopore. By using latent-track membranes with diameter of 0.6 nm, we found the channels are nearly non-conductive in small voltage due to the blockade of cations bound at surface, however turns to be conductive as rising of voltage due to releasing of bound ions, which differs from the mechanisms in $MoS_2$ nanopore. By Kramers’ escape framework, we rationalized an analytical equation to fit experimental results, uncovering new fundamental insights of ion transport in the smallest channels.

\end{abstract}

\maketitle

\paragraph*{Introduction--} 
Ionic transport in an angstrom scale channel is critical to understand and mimic the physiological mechanism of biology ion channels\cite{faucher_critical_2019,bocquet_nanofluidics_2020,robin_modeling_2021}. The developing of nanotechnology and materials enable to create angstrom-scale structures to study ion transport in such confinement, which has led to several peculiar discoveries\cite{radha_molecular_2016,fumagalli_anomalously_2018,bocquet_nanofluidics_2020,kavokine_ionic_2019} and excellent separation capabilities\cite{sun_exponentially_2021,feng_single-layer_2016,fu_dehydration-determined_2020}. The Coulomb Blockade (CB) is a unique phenomenon discovered in angstrom-scale channels \cite{feng_single-layer_2016,kavokine_fluids_2021}, which originated from solid nanoelectronics devices that electrons must overcome an energy barrier to transport through an island of electrons\cite{fulton_observation_1987,clarke_transport_1997}. The ionic CB in nanofluidics were proposed by M. Di Ventra et al., in MD simulation\cite{krems_ionic_2013}, which was considered as one of the possible reasons of ion selectivity in biology ion channels\cite{kaufman_coulomb_2015,chernev_prospects_2020,kaufman_multi-ion_2013}. Feng et.al discovered a non-linear conduction by using nanopores on a single layer $MoS_2$, attributed to the joint action of ion-dehydration and self-energy barrier as the ionic CB effects\cite{feng_single-layer_2016}. Although theories progressed rapidly including mechanisms of dehydration \cite{barabash_origin_2021,fu_dehydration-determined_2020}, self-energy\cite{parsegian_energy_1969,feng_single-layer_2016,teber_translocation_2005}, Wien-effects\cite{kavokine_ionic_2019,kavokine_interaction_2022,coquinot_quantum_2022}, 
in addition more types of  angstrom-scale pores/materials were used in the study of ionic transport\cite{cheng_molecular_2021,jain_heterogeneous_2015,lu_ultrafast_2022, suk_ion_2014}, the CB effects in experiments were never reported in artificial channels since it was discovered in $MoS_2$ nanopore\cite{chernev_prospects_2020}. 

In this work, we fabricated the latent ion track channels with a simplified procedure according to the previous work \cite{wen_highly_2016,wang_ultrafast_2018}. We characterized the channels with a most probable diameter of 0.60nm by isotherm adsorption of $CO_2$ gas molecules using Brunauer-Emmett-Teller (BET) model \cite{brunauer_adsorption_1938,chen_synergistic_2019}, quantitatively approving the angstrom-scale latent track channels which were indeed a challenge. 

\begin{figure}[b]
    \centering
    \setcounter {subfigure} {0}(a){
		\includegraphics[width=0.2\textwidth]{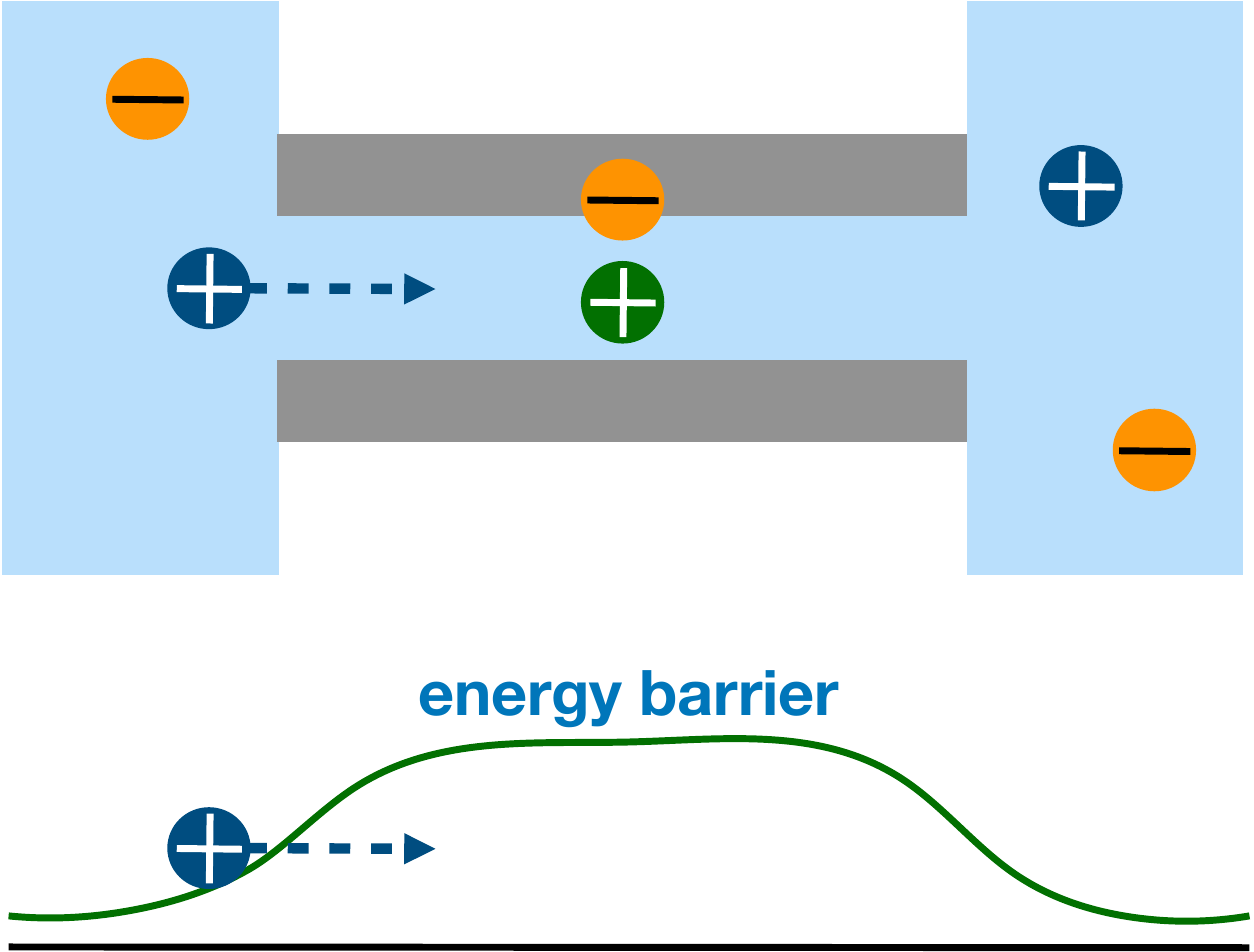}}
	\setcounter {subfigure} {0}(b){
		\includegraphics[width=0.2\textwidth]{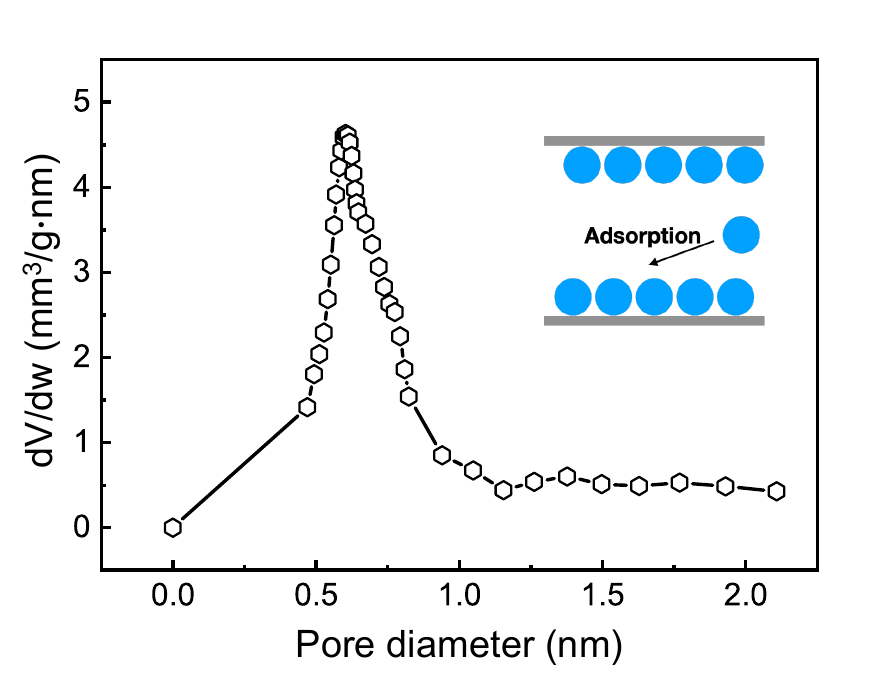}}
\caption{(a). The schematic picture of CB effect. A cation strongly bound to the surface charge, blockade the ion transport through the angstrom-scale channel. (b). The diameter distribution of latent track channels, characterized by the isotherm adsorption of $CO_2$ gas molecules. Our results showed the most probable diameter of latent track channels was 0.60nm.} 
\label{fig:1}
\end{figure}

\begin{figure*} [htp]
	\centering
   \setcounter {subfigure} {1} (a){
		\includegraphics[width=0.28\textwidth]{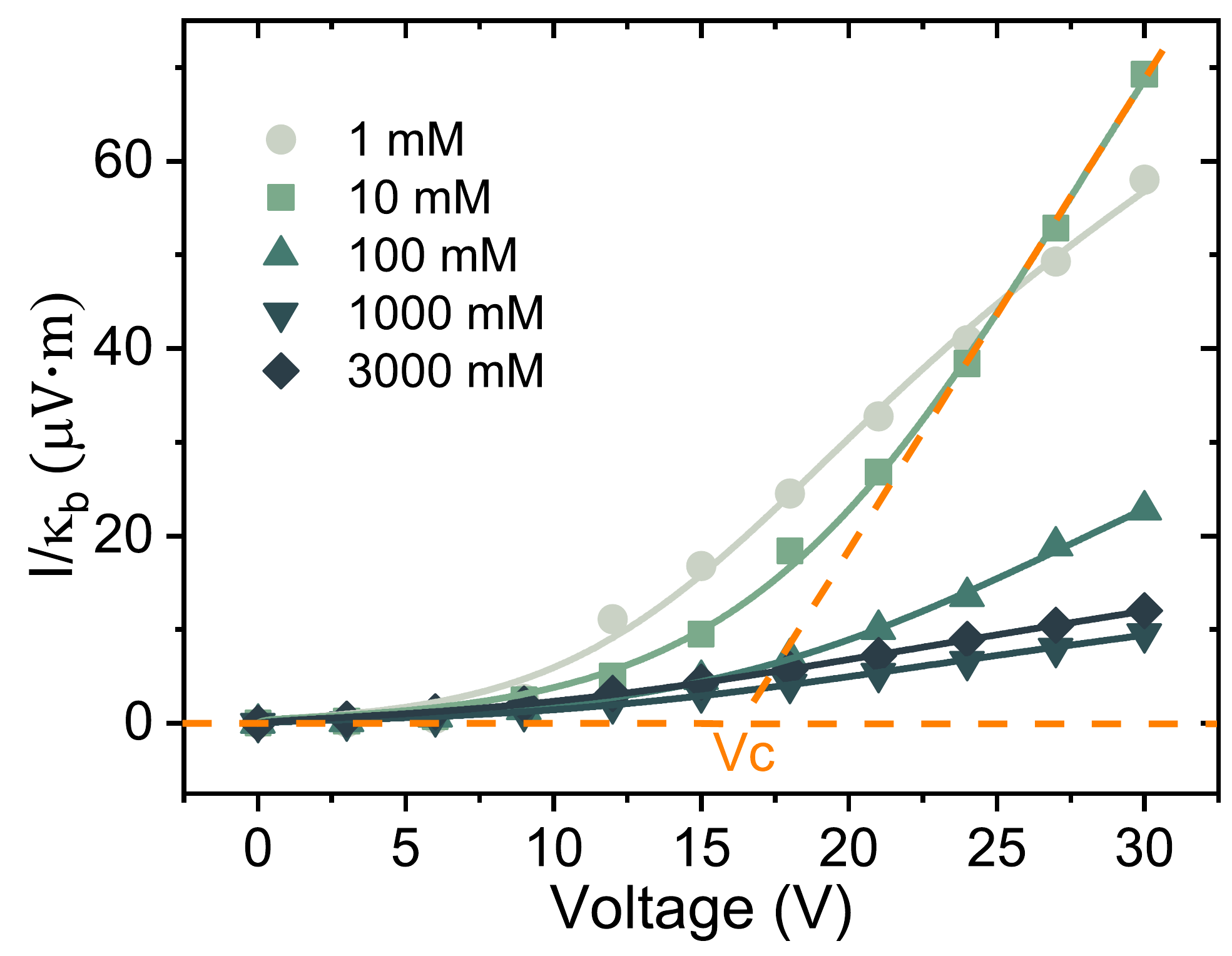}}
    \setcounter {subfigure} {1} (b){
		\includegraphics[width=0.28\textwidth]{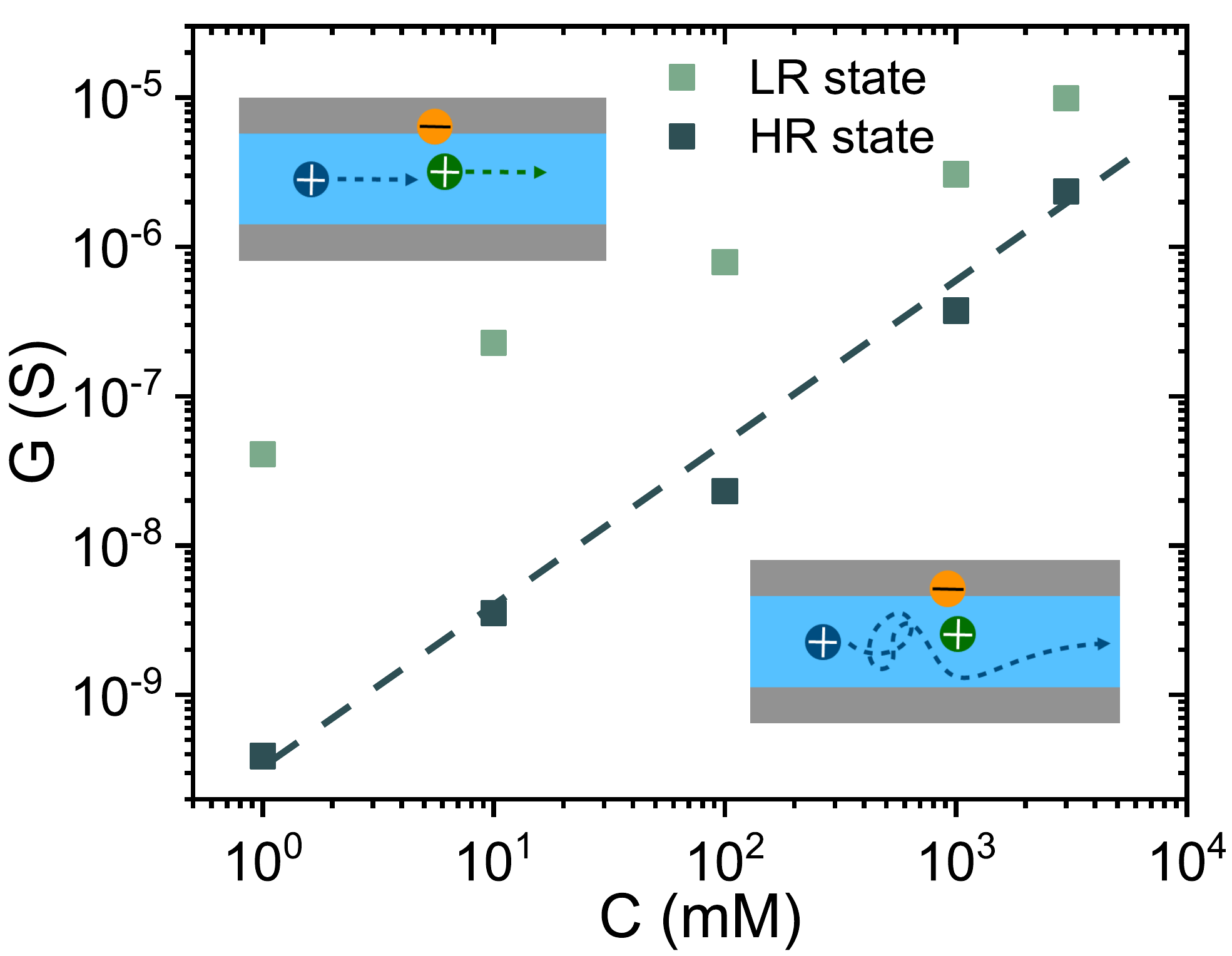}}
    \setcounter {subfigure} {1} (c){
		\includegraphics[width=0.28\textwidth]{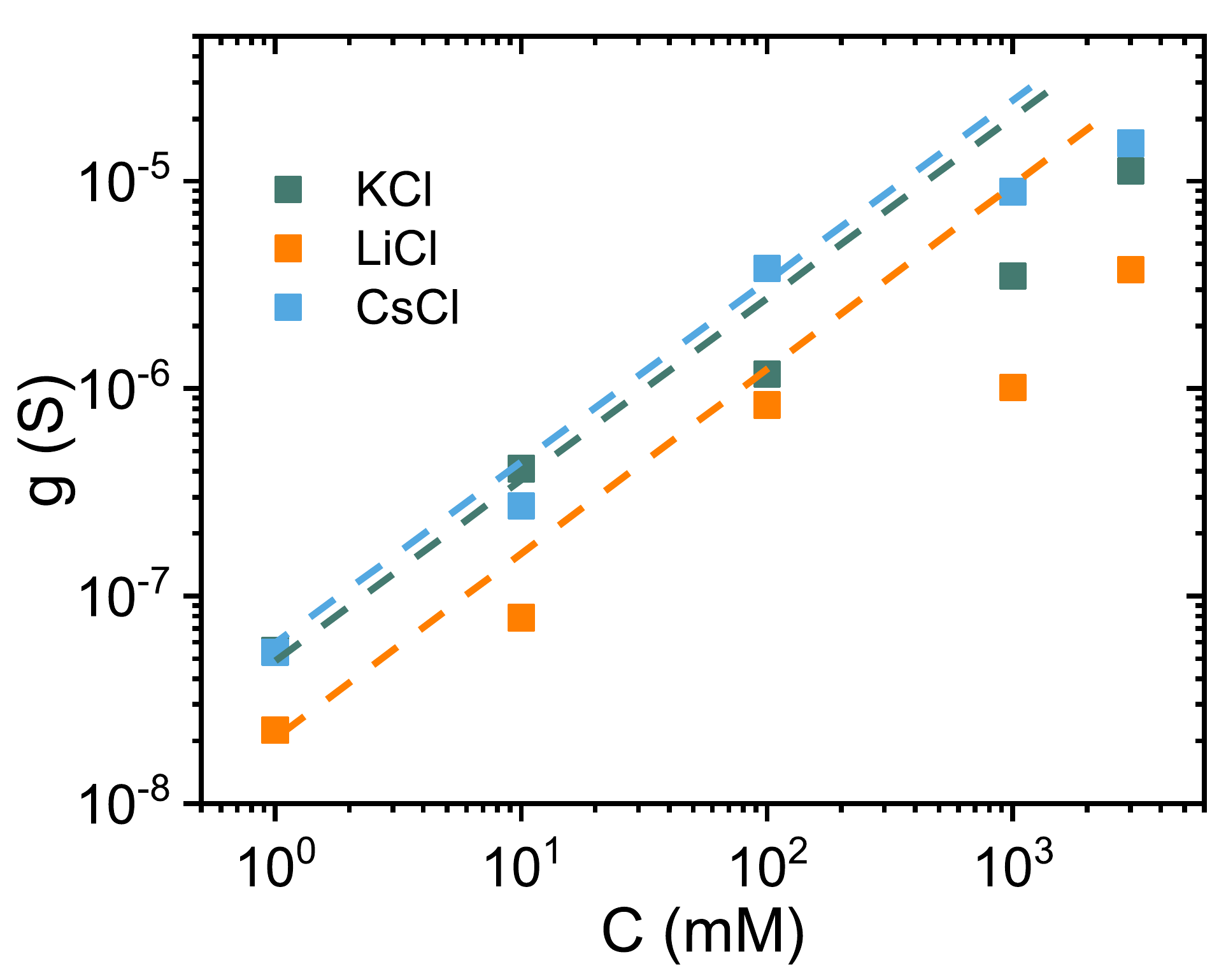}}

\caption{the typical I-V curves at positive applied potential in various concentrations. (b) The conductance of the HR (deep green) and LR (light green) as a function of concentration respectively. The possible mechanism of ionic transport were schematically drawn as inset figures. (c) the fitted conduction of $Li^+$, $K^+$, $Cs^+$ agreed well with the slope of $\kappa_b$ when $C < 0.1M$, while apparently decreased when $C > 1M$. } 
\label{fig:2}
\end{figure*}

We found a transition of high resistance (HR) to low resistance (LR) of the membranes as rising of an external electrical field in diluted salt solutions. Although the non-linear conduction in latent track channels appeared in the previous work\cite{wen_highly_2016,wu_preparation_2021,cheng_ionic_2021}, the mechanisms and analysis of the results were still far from clear. We found the power law of conductance at HR stage matched well with the conductivity of bulk solution, implying the conduction at low voltage is independent of the surface charge density. Besides, the conduction at HR state suggests only effectively $\sim 0.1\%$ of channels were conductive by radius of 0.3nm. We suspect it was caused by the counterions bound to the surface charge, which blockades the transport of free ions (scheme shown in Fig \ref{fig:1}a), approval by our MD simulation and well explained the mechanisms of conductance transition. Our CB mechanism differs from the CB effects discovered in single layer $MoS_2$ caused by self-energy barrier and dehydration of ions \cite{feng_single-layer_2016}. 

Using Kramer’s escape framework, we rationalized with an analytical equation of the current to fit our experimental results. The fitted values uncovers insightful details of the ionic transport in angstrom scale channels, such as decreases of conductivity at concentrated solution, and reduction of energy barrier as rising of ion density and surface charge density in the channel, resulting a decrease of threshold voltage transiting to conductive state. Our results matched well with the Coulomb Blockade effects predicted by Kavokine \cite{kavokine_interaction_2022}, at a dielectric surface of “infinite” long angstrom-scale channels. Our experimental results and theoretical analysis possibly give insights of the ion transport in well-defined 1D angstrom-scale channel, useful for fundamental studies of ionic transport or practical applications in angstrom scale channels.

\paragraph*{Fabrications and Experimental setup--} 

We first irradiated the polymer film by Kr ion with energy of over 1 MeV/u and density of $3 \times 10^8 cm^{-2}$ to form latent tracks in PET polymer films, from Lanzhou Heavy Ion Research Facility (HIRFL). Then we placed specimens under UV (365 nm) exposure with the illumination density of $65 mW\cdot cm^{-2}$ for over 30 minutes at each side. Finally, we clamped the irradiated polymer film in the 1mM KCl electrolyte solution bath in 50 degrees under voltage scanning ($\pm$10V), to remove the products of radiolysis in the latent tracks thus fabricating one-dimensional ultrafine channels. The voltage scanning lasts for about half an hour until the amplitude of current get saturated shown as Fig.S1b. The film will be carefully cleaned by filling DI water under sweeping voltage, to remove the residual salts in the channels. The films were well preserved and dried in the air for overnight. For each experimental measurement, we keep the similar cleaning procedure above. 

Precisely characterizing the size of the latent track channels is indeed a challenge. Traditional methods, including resistance measurements in 1M KCl, are inapplicable because it does not obey the classical conduction laws, as we demonstrated later. Hereby we use isotherms adsorption of $CO_2$ gas to characterize the size of channel, with details as follows.

To maximize surface area on the film, we irradiated the PET film with density of $10^{11} cm^{-2}$ (6$\mu$m thick) finally obtained $2.9 m^2/g$ BET surface area measured by Micromeritics 3Flex. We placed specimens in a vacuumed chamber and gradually pressurized with $CO_2$ gas as rising of pressure of $P/P_0$ where $P_0$ is the atmosphere pressure. We measured the volume of absorbent in each applied pressure by recording the volume change of $CO_2$ over 15 seconds as a time step of equilibrium (see SM). As a result, we could linearly fit the adsorption kinetics according to Langmuir equation and calculate the surface area of the film. Finally, we calculated the diameter of channel distributed from 0.46 to 1.0 nm with a most probable diameter of 0.60 nm, where the smaller size were not detectable as the $CO_2$ molecules may not forming uniform Langmuir adsorption in such small channels (Fig \ref{fig:1}b). However, the factors during irradiation energy \cite{wang_saxs_2022,apel_influence_2009}, UV exposure \cite{zhu_influence_2005}, procedure of removing radiolysis products \cite{apel_soft-etched_2022,wang_ultrafast_2018} on the channel size are still not clear, even not the applications \cite{ali_haider_osmotic_2022} . 

\paragraph*{Results and Discussion--} 
We first investigated the conductance as a function of the salt concentration $C$, ranging from 1 mM to 3M KCl at pH 5.0. To well present the current response in various concentrations, we showed the current divided by bulk conductivity of solution $I/\kappa_b$ instead of current I in Fig \ref{fig:2}a. We found non-linear I-V curves in low $C$. It appears a high resistance (HR) state at amplitude of voltage smaller than $\sim 3V$ for both polarity of electrical fields (See SM). Here we only demonstrated the results in positive bias voltages as examples for the analysis. 

For the HR state, we linear fits the I-V curves at voltage smaller than 3V to obtain the conductance $G_{HR}$ (dark green dots in Figure \ref{fig:2}b), and calculated $G_{LR}=I_{+30V}/V_{+30V}$ as the conductance of LR state (light green dots in Figure \ref{fig:2} b). We found the $G_{HR}$ perfectly matched with the power-law of $\kappa_b$ (dashed lines) without a plateau in low concentrations as classical nanofluidics\cite{stein_surface-charge-governed_2004}, which possibly implies the conduction is independent of surface charges under such low voltage. We calculated the numbers of conductive channels by $G_{HR}$ using classical nanofluidic conduction theories which shows only $\sim 0.1\%$ channels are conductive using diameter of 0.60nm. To validate the electrical field dependent conduction, we investigated conduction using thicknesses of $2.5 \mu m$, $6 \mu m$, and $12 \mu m$. The results show that the threshold voltage for conductance transition decreases with the film thickness, but remains nearly a constant at $ \sim 1V/\mu m$. The results are available in SM.

Above clues from experiments inspired us that the HR states of our results were possibly caused by the ionic Coulomb blockade. We performed MD simulations in the CNT with effective diameter of $0.55nm$ comparable to the diameter measured by isotherm adsorption (see SM). We set a single carbon atom at surface been fully charged by an elementary charge of $e$, to mimic the dissociation of carboxyl groups at dielectric surface. Similar as previous work \cite{kavokine_ionic_2019,qiao_atypical_2003,xie_liquid-solid_2020}, we found strong Coulomb interaction between the surface charge and counterions, resulting the blockade of ionic transport through the angstrom-scale channel. The small electrical field is not able to “excite” the bound ion, while the conduction relies on the free ions passing by the bound ions. As the electrical field increases, the probability of releasing the bound ions increases, resulting an "open" state for ion transport as a LR state. 

Besides the Coulomb interaction between cation and surface charge, additional energy barrier including dehydration \cite{fu_dehydration-determined_2020,sahu_dehydration_2017}, self-energy \cite{parsegian_energy_1969}, and Bjerrum ion pairs \cite{kavokine_ionic_2019} may play a role. For an approximation, we considered an equivalent parabolic potential barrier with a height of potential barrier $\Delta U$, which may from one or joint actions of above effects. Thus, we have an analytical expression for the 1D ionic transport by Kramer’s escape approximation. The detailed derivation of current can be found in the SM. Thus, we have the following form of equations that used for fitting I-V curves, slightly different from the Hodgkin-Huxley model  \cite{hodgkin_quantitative_1952}.

\begin{equation}
\label{eq:I}
I=\frac{g V}{1+ k V e^{-b V}}
\end{equation}

\begin{figure}[t]
\centering
\setcounter {subfigure} {2} (a){
		\includegraphics[width=0.3\textwidth]{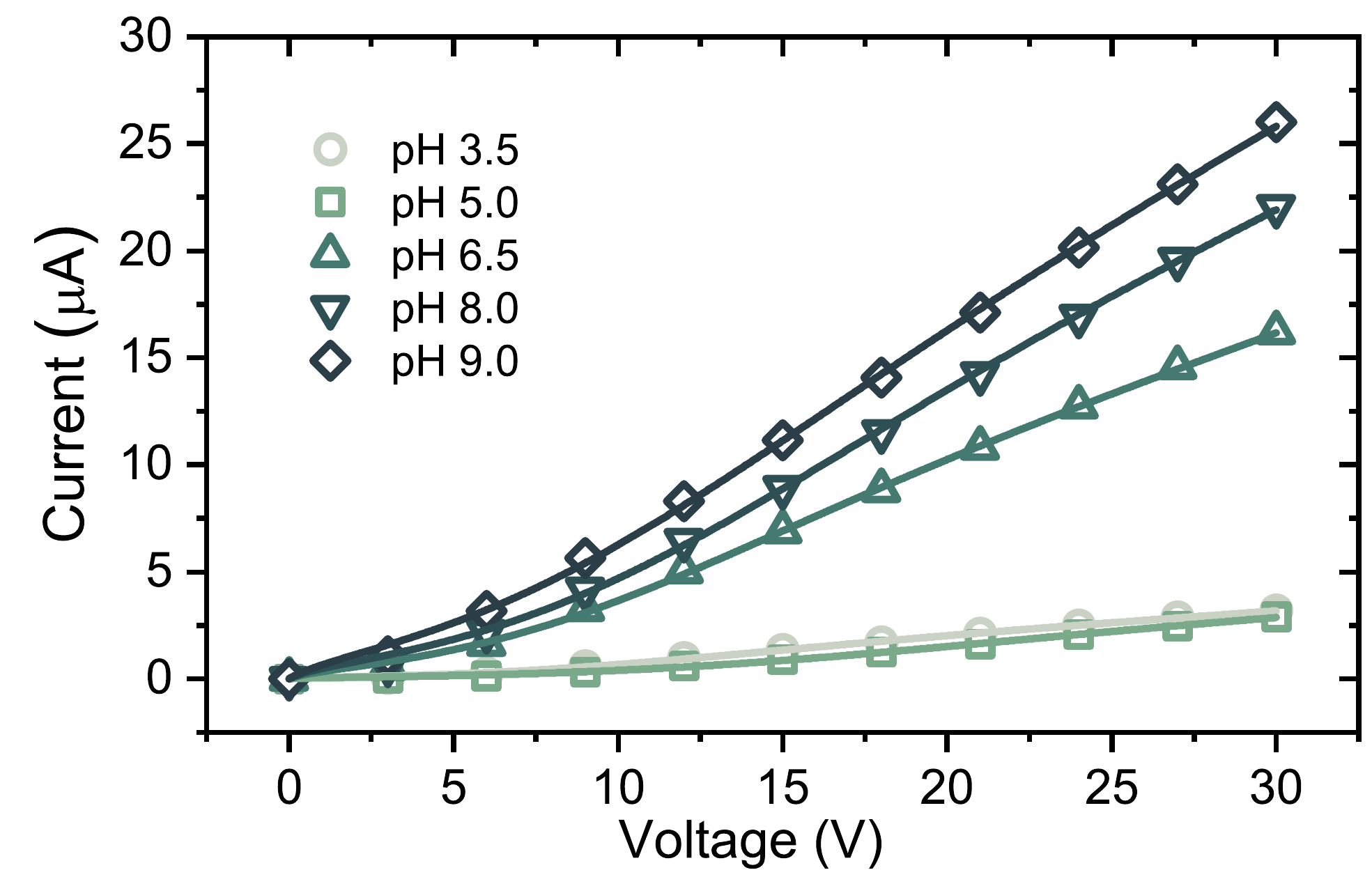}}
        \\
    \setcounter {subfigure} {2} (b){
		\includegraphics[width=0.3\textwidth]{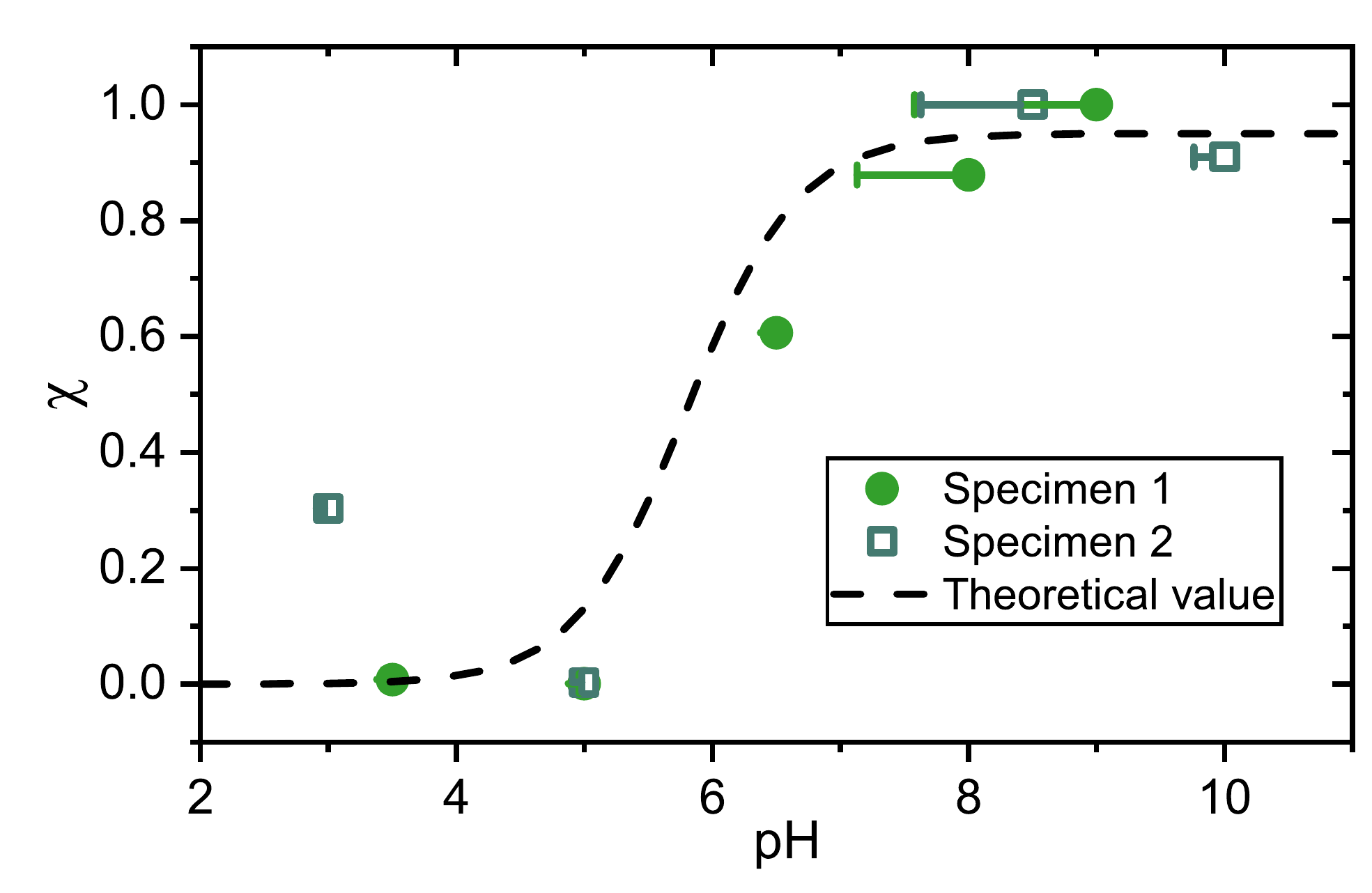}}
\caption{(a) the typical I-V curves in 10 mM KCl solution with pH from 3.5 to 9.5. (b) the fitted conductance from the individual experiments, matched well with the chemical equilibrium in different pH values.} 
\label{fig:3}
\end{figure}

where the $g, k$ and $b$ are the fitted parameters, and $V$ is the applied voltage. The fitting parameter $g = N\cdot \mu CA_{ch}/L$ are the conduction of all channels with number of $N$, considering free ions from the bulk solution with concentration of $C_b$ and bound ions $C_s$ at surface $C=C_b+C_s$. The $A_{ch}$ is the mean cross-sectional area of a single channel. When energy barrier is negligible, the equation \ref{eq:I} turns to the classical nanofluidic conduction that linear increases as the applied voltage \cite{bocquet_nanofluidics_2010}. For the energy barrier dominated ion transport, it becomes the same formula as Arrhenius type behavior in biology ion channels \cite{chernyshev_thermodynamic_2002} and angstrom-scale slits \cite{hu_transport_2018,sun_exponentially_2021}. The entrance effects was neglected as we have “infinite” long channels ($r<<L$) \cite{ma_entrance_2018,macha_2d_2019}. Besides, we only considered the transport of metal ions as we found the $Cl^-$ were not involved in the conduction in negatively charged angstrom channels in MD simulation. The fitting parameter $k \sim \mu  e^{\Delta U/{k_B T}}$ where the $\Delta U$ is the energy barrier for ion transport through the confined channel. The fitting parameter $b$ indicates the strength of electrical force on the bound ions, which is a constant between 0.20 to 0.25 $V^{-1}$ for each specimen as in different concentrations (See SM), however slightly rises to maximal 0.4 $V^{-1}$ as pH increases. The deviation of $b$ in different specimens is possibly relevant to fabrication procedures such as radiolysis of films, which still need to be studied in further. 

The fitted conduction $g$ was shown as solid dots in Fig \ref{fig:2}c as a function of $C_b$ with different monovalent salt solution. The dashed lines represent the slope of $\kappa_b$ in various salt solutions. The power law of experimental conduction (solid dots) matched well with $\kappa_b$ below 0.1M, however obviously getting smaller than $\kappa_b$ above 1M. We suspect this reduction of conductivity was caused by the formation of ion pairs in the angstrom scale channel, previously reported in the slits nanochannels \cite{zhao_two-dimensional_2021,kavokine_ionic_2019} and CNT  \cite{nicholson_ion_2003,kavokine_ionic_2019,neklyudov_putting_2022,aydin_ion_2021}. 
Since only free ions contribute to the conduction in small HR state, we suspect the formation of ion pairs as increases of $C_b$ results the reduction of conductivity\cite{yoon_electrical_2019}.

Then we studied the conduction under various pH solutions (Fig.\ref{fig:3}a). Our results showed the amplitude of conduction gradually increases as the pH values. In addition, the system gets close to Ohmic as the pH rises. The fitted conduction $g$ including the bound ions in the channel gradually increases and then get saturated as the pH values, shown in the Fig. \ref{fig:3}b. According to the chemical equilibrium at surface that the surface charge density varies as the pH using following equations \cite{behrens_charge_2001}. 

\begin{equation}
\label{eq:sigma}
\Sigma = - e \Gamma \frac{10^{-pK}}{10^{-pK}+10^{-pH}}
\end{equation}
where $e$ is the elementary charge. We obtained the site density $\Gamma$ of carboxyl groups at the surface as $\sim 10^{-2} site/nm^2$ that is smaller than that in the track etched nanopores \cite{lin_voltage-induced_2018}, possibly due to the lacking of chemical etching. We obtained dissociation factor as $\chi = 10^{-pK}/(10^{-pK}+10^{-pH})$, where the $pK$ was taken as 5.8 for the carboxyl group \cite{behrens_charge_2001}. Taking the counterions in the conductance $g_\Sigma =N\cdot \mu C_\Sigma A_{ch}/L$ where $C_\Sigma = \frac{2 \Gamma}{r N_A}$ and $N_A$ is Avogadro number, we have a theoretical value shown as solid line in Fig. \ref{fig:3} b. The buffer solutions were avoided due to their profound effects for chemical equilibrium at surface \cite{bostrom_specific_2004}. The error bars indicate the pH change during the measurements ($\sim$ 20 mins), caused by the adsorption of $CO_2$ from air. The deviation of experimental data at pH=3.0 possible caused by the inversion of surface charge, which doesn't fits the theoretical prediction in equation \ref{eq:sigma}.

\begin{figure}[t]
\centering
\setcounter {subfigure} {3} (a){
		\includegraphics[width=0.3\textwidth]{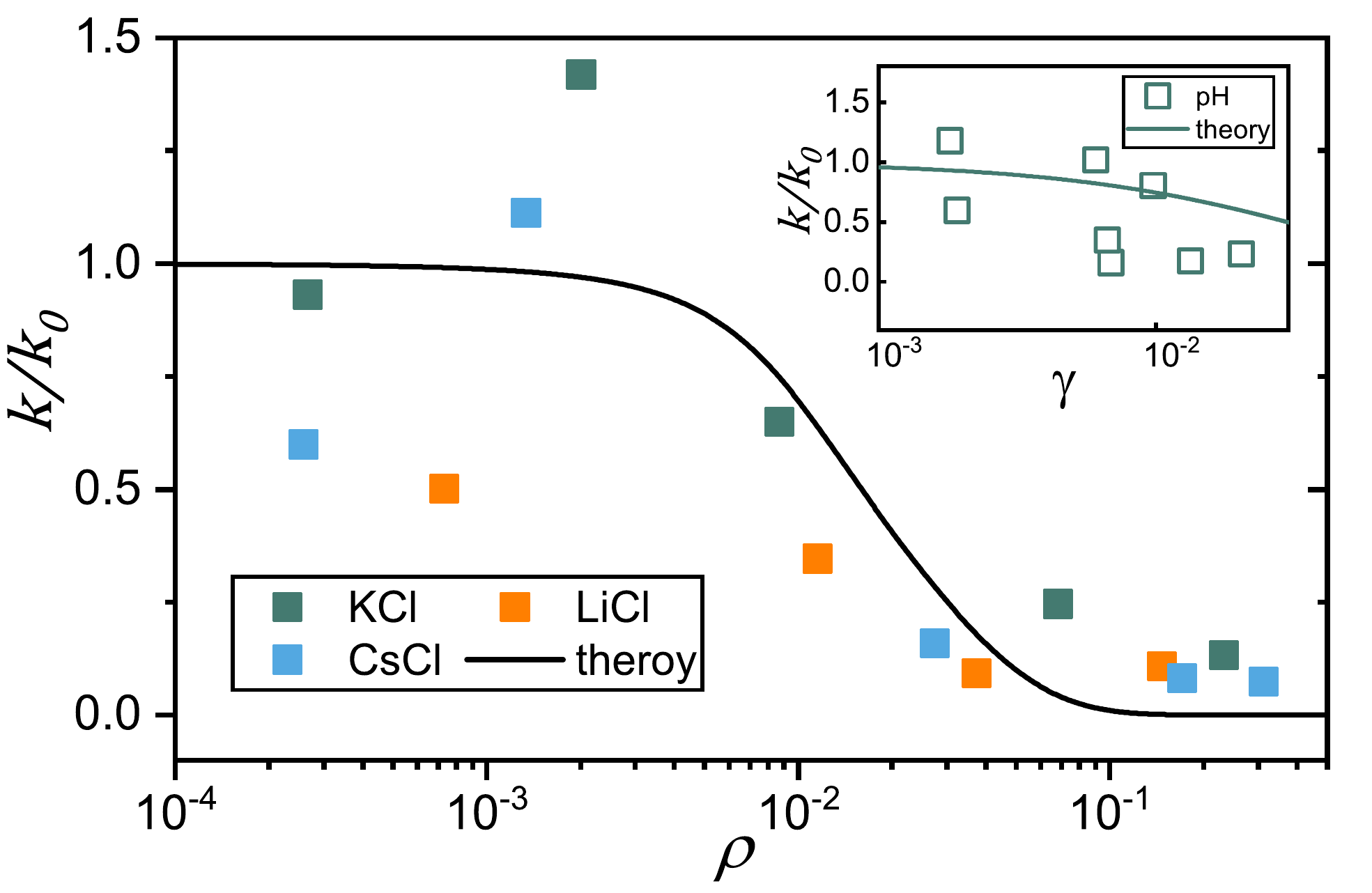}}
        \\
    \setcounter {subfigure} {3} (b){
		\includegraphics[width=0.3\textwidth]{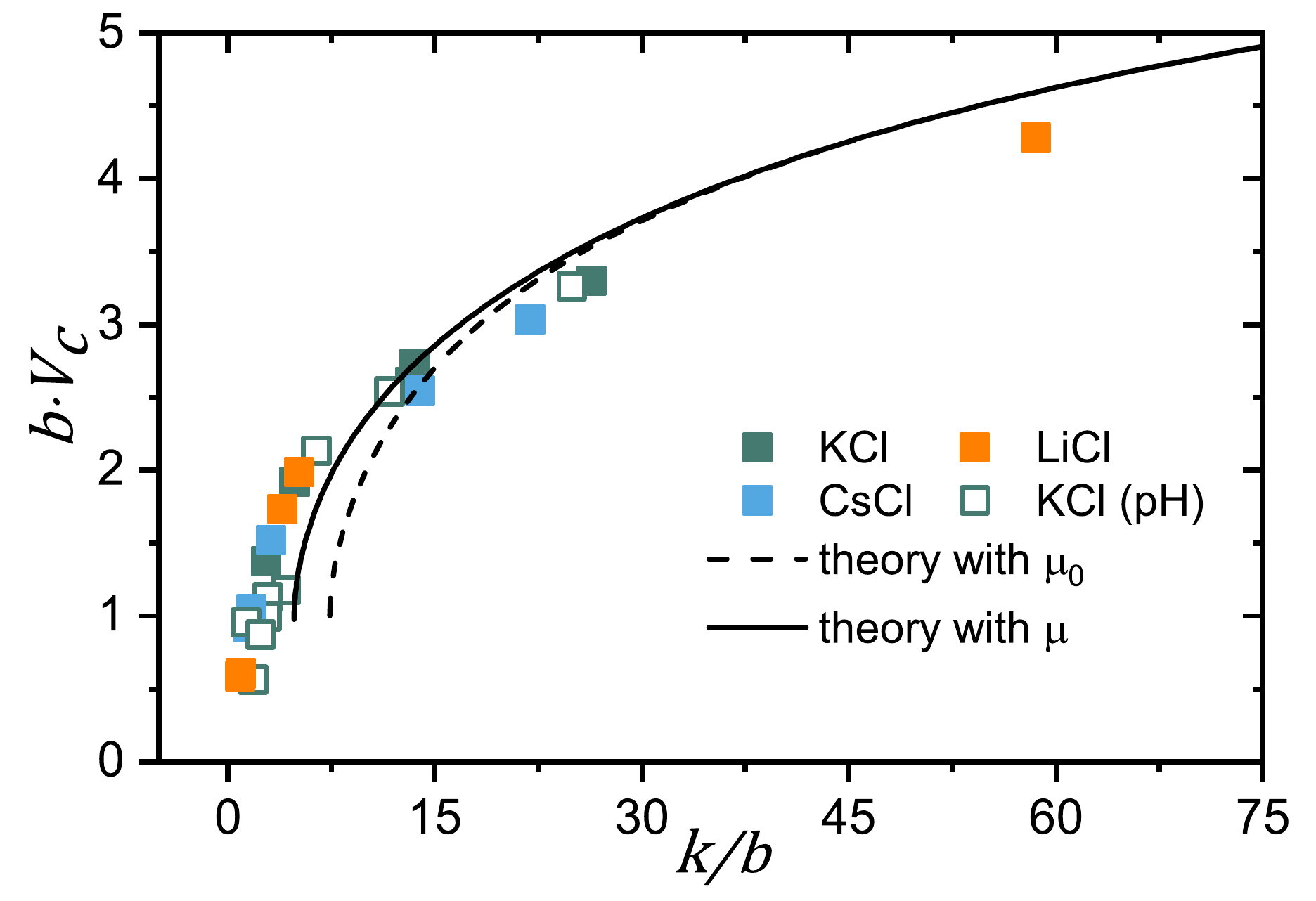}}
\caption{(a) The fitted parameter $k/k_0$ decreases as $\rho$ and $\gamma$ (inset) matching well with the theoretical predictions (solid line). (b) the critical voltage increases as $k/\mu$, matching well with the experiments calculated with constant ion mobility (dashed line) and in mobility decreases in high concentrations(solid line).} 
\label{fig:4}
\end{figure}

Now we focused another fitting parameter $k$. The linear density of free ions can be calculated as $d_f=(A_{ch}N_A C_b)^{-1} $, while the linear density of bound ion density as $d_s=(A_{ch}N_A\frac{2}{r} \frac{\Sigma}{e})^{-1}$. We have the linear ion density $\rho=x_T/d_f$, where $x_T=a^2 /( 2 l_B )$ is thermal length of an ion. The potential barrier $U$ rapidly decreases as the entropy of the channel increases due to the dissociation of bound ion or ion pairs predicted by Zhang as follows \cite{zhang_conductance_2005,zhang_ion_2006} . 

\begin{equation}
\label{eq:k}
U = U_0 [1- 4 \gamma ln(\frac{1}{2 \rho} \sinh(\frac{\rho}{\gamma}))]
\end{equation}

Finally, we evaluated the $k/k_0 \sim \mu e^{(U-U_0)/{k_B T}}$ as a function of ion density $\rho$ and surface charge density $\gamma$ shown as a solid line in Fig.\ref{fig:4}a, considering the reduction of mobility compared to the bulk values. The increases of ion density, including the concentration of free ions and bound ions at surface, reducing the energy barrier of transport. As a result, current getting close to the linear response to the applied voltage. We statistically counted all fitted $k/k_0$ as a function of $\rho$ derived from $C_b$ shown in \ref{fig:4}a, and $\gamma = x_T \cdot 2\pi r \Gamma \chi$ derived from the equation \ref{eq:sigma} as an inset figure in \ref{fig:4}a. The theoretical $k/k_0$ by equation \ref{eq:k} matched well with the fitted values in experiments. 

To quantitatively evaluate the critical voltage of conduction states transition, we linear fits the I-V curves at LR states and considered a crossing point to the X-axis as $V_C$, with an example shown as a dashed line in the I-V curves in Fig.\ref{fig:2}a. Details of deriving the critical voltage in experiments can be found in the SM. From the theoretical aspects, we estimate the transition voltage $V_C$ by $k Ve^{-bV_C} = e$, where the current starts rising apparently. Thus we have the solution as follows, where the $W(z)$ is Lambert W Function. 
\begin{equation}
\label{eq:Vc}
V_C = -W(-\frac{e \cdot b}{k})/b
\end{equation}
With all fitted $k$ and $b$ from experiments, we showed the $V_C$ considering different types of salts, concentrations and pH values of the film in the Fig. \ref{fig:4}b. The theoretical values by equation \ref{eq:Vc} considering the free ion density $\rho$ and surface charge density $\gamma$, well predicting the tendency of $V_C$ in experiments that is helpful for understanding 1D ionic transport due to CB effects. 

\paragraph*{Conclusion--} 
In conclusion, we fabricated latent ion-track channels and characterized 0.60nm as most probable diameter by the isotherm gas adsorptions. We found a resistance transition from HR to LR state as rises of applied voltage in diluted solutions. We suspect the HR state was caused by the counterions bound to the surface charge that blockades the ion transport due to the strong Coulomb interaction to the surface charge in confinements, approved by MD simulations. The conduction at HR state matched well with the slope of bulk conductivity, illustrating the counterions didn't contribute to the conduction. As increases of $E$, the probability of releasing bound ions increases, resulting LR state. We rationalized an analytical equation by Kramer’s escaping approximation, with an equivalent parabolic potential barrier for ion transport, uncovering more insights of ion transport, such as the reduction of conductivity in channels at concentrated salt solutions and decrease of energy barrier as increases of ions density. Our results will be useful for the fundamental studies of ion transport in the smallest scale channels, as well as the application aspects like separations. 

\begin{acknowledgments}
The authors thank Jinglai Duan, Jie Liu and Guanghua Du in HIRFL for the help of film irradiation, Shusong Zhang for useful discussions, Xianzhi Ke and Shenghui Guo for the experimental help, Jianwei Cao and Kaijie Chen for the help of isotherm adsorption. This work is supported by NSFC No. 12075191. 
\end{acknowledgments}

\nocite{*}

\bibliography{angstrom-transport_WWH}

\end{document}